\documentclass[manuscript, screen]{acmart}
\AtBeginDocument{%
  }

\setcopyright{acmlicensed}
\copyrightyear{2018}
\acmYear{2018}
\acmDOI{XXXXXXX.XXXXXXX}
\acmConference[Conference acronym 'XX]{Make sure to enter the correct
  conference title from your rights confirmation email}{June 03--05,
  2018}{Woodstock, NY}
\acmISBN{978-1-4503-XXXX-X/2018/06}

\begin{document}

\title{AdaPT: Adaptive Lesson Plan Transformer for Cross-Regional and Differentiated Instruction}


\author{Yanjie ZHANG}
\affiliation{%
  \institution{The Hong Kong University of Science and Technology}
  \city{Hong Kong}
  \country{China}}
\email{yzhangvj@connect.ust.hk}

\author{Jiajun Zhu}
\affiliation{%
  \institution{Zhejiang University}
  \city{Hangzhou}
  \country{China}
}
\email{jiajunzhuchris@gmail.com}

\author{Minyu Wu*}
\affiliation{%
 \institution{Shanghai Normal University}
 \city{Shanghai}
 \country{China}}
\email{wuminyu@shnu.edu.cn}

\author{Huamin Qu}
\affiliation{%
  \institution{The Hong Kong University of Science and Technology}
  \city{Hong Kong}
  \country{China}}
\email{huamin@ust.hk}

\author{Sicheng Song*}
\affiliation{%
  \department{School of Data Science and Engineering}
  \institution{East China Normal University}
  \city{Shanghai}
  \country{China}}
\email{scsong@dase.ecnu.edu.cn}

\renewcommand{\shortauthors}{Trovato et al.}

\begin{abstract}
Due to educational inequality, high-quality lesson plans often mismatch the needs of disparate educational contexts. Teachers typically modify existing lesson plans to fit new contexts, but current tools instead focus on generating content from scratch, creating additional workload. Moreover, a critical gap remains in supporting teachers to quickly adapt to new learning profiles. To bridge these gaps, we present AdaPT, a system leverages LLMs to support transformation of existing lesson plans for cross-regional and differentiated instruction. AdaPT features an interactive interface that allows teachers to input student profiles, offers structured lesson representation, provides explanations for lesson-plan transformations, automatically adapts lesson content for new contexts, and supports iterative, teacher-in-the-loop refinement. We evaluated AdaPT through a user study with 9 teachers and an expert evaluation with 3 specialists. Results show that AdaPT supports workflows of teachers and offers a promising pathway toward promoting educational equity.
\end{abstract}


\begin{teaserfigure}
  \includegraphics[width=\textwidth]{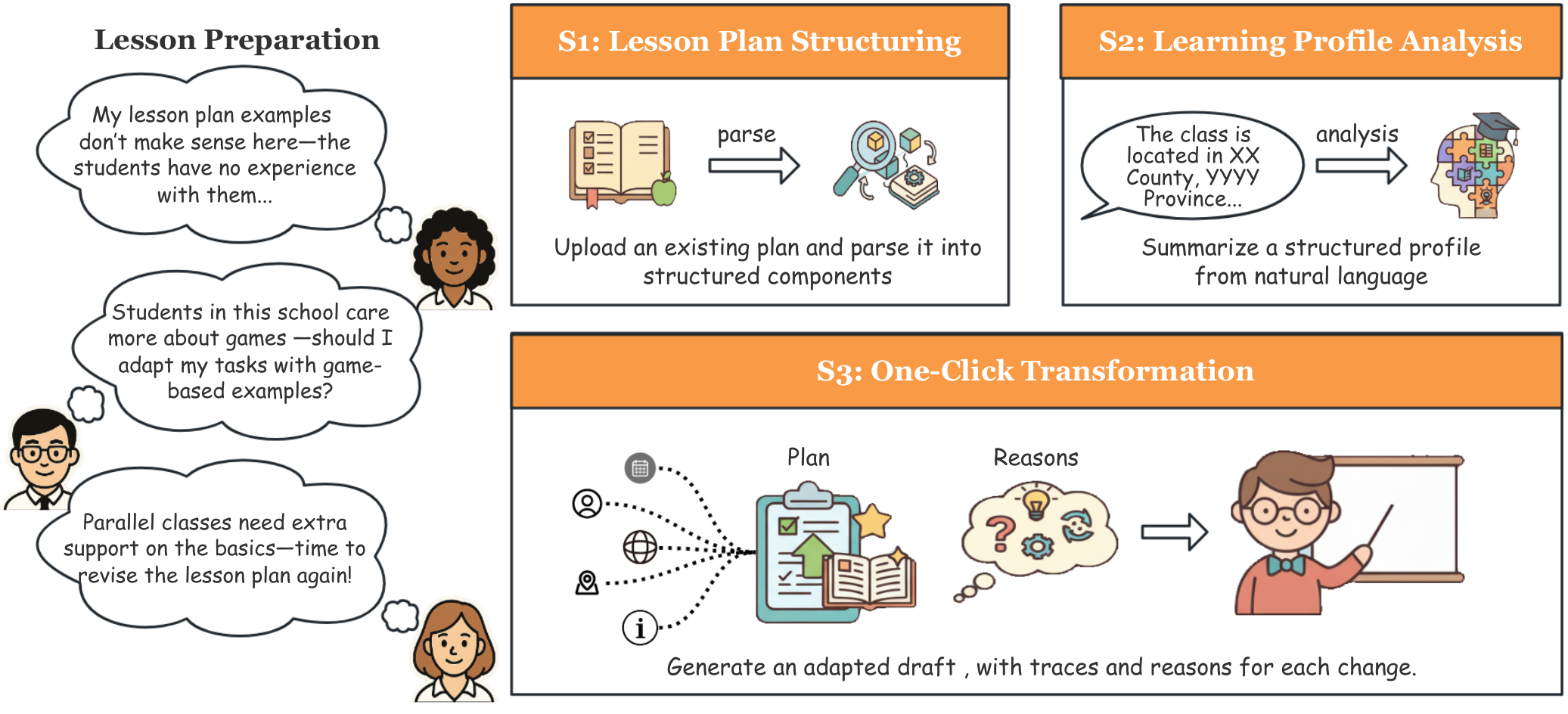}
  \caption{AdaPT supports (S1) lesson plan structuring, (S2) learning profile analysis, and (S3) one-click transformation, enabling efficient and context-aware adaptation of existing lesson plans. More details are presented in Section~4.}
  \Description{The teaser figure illustrates teachers’ challenges in lesson plan adaptation (left) and the AdaPT system workflow (right). The workflow includes lesson plan structuring, learning profile analysis, and one-click transformation to produce adapted lesson plans.}
  \label{fig:teaser}
\end{teaserfigure}

\received{20 February 2007}
\received[revised]{12 March 2009}
\received[accepted]{5 June 2009}

\maketitle

\section{Introduction}
Educational inequality remains one of the most pressing challenges in China, manifesting across regions (eastern vs. western, urban vs. rural)\cite{hannum2007education, zhang2009spatial}, across schools within the same region\cite{long2020educational}, and across classes within the same school\cite{gupta2023association}. 
Such disparities lead to significant differences in students’ access to resources, learning experiences, and academic achievement.
Prior work in educational research has documented that students in rural or under-resourced areas often struggle not because of lower motivation, but due to the lack of appropriate instructional resources and contextualized examples \cite{irvin2012educational}. 
A promising strategy to mitigate such inequality lies in adapting lesson content to local contexts, thereby ensuring that diverse groups of students can engage with instructional materials that are both relatable and accessible\cite{lin2024quest, zhang2023context}.
Teachers have long relied on differentiated instruction to tailor lessons to diverse student profiles.
For example, high-performing classes may be assigned more advanced, creative tasks, while lower-performing classes focus on foundational knowledge. In rural settings, teachers often substitute urban-centered examples with those drawn from students’ immediate environments. 
Meanwhile, ICT-mediated approaches—such as live-streaming classrooms (\textit{One Screen} in China \cite{neteaseOneScreen}) and online resource platforms—have been increasingly deployed to reduce regional disparities\cite{smartedu2025}. 
Although promising, these interventions often impose substantial additional workload on teachers, who must manually revise lesson plans for different classes or contexts. 
Without systematic support, such practices are difficult to sustain at scale. 
This tension underscores the need for tools that can intelligently adapt lesson content to varied teaching conditions, thereby alleviating teachers’ preparation burden while promoting equity.

The rapid advancement of large language models (LLMs) offers new possibilities for addressing these challenges. Recent studies have begun to explore LLMs in educational contexts, such as generating learning content, lesson plans, and analogies to aid student comprehension. 
For example, LessonPlanner~\cite{fan2024lessonplanner} demonstrated that LLMs can scaffold lesson design by producing outlines and suggested activities. Shao et al.~\cite{shao2025unlocking} further showed that LLM-generated analogies can help students understand abstract scientific concepts by relating them to familiar domains. Collectively, these works suggest that LLMs can augment teachers’ work by offering flexible, in-situ content generation. However, existing systems primarily emphasize generating lesson plans from scratch. In practice, teachers rarely create entirely new lesson plans for each setting; instead, they routinely adapt and transform existing ones. This reveals a critical gap in current research: how to design LLM-powered systems that directly support lesson plan transformation—adapting existing teaching materials to diverse contexts and student needs.

To understand teachers’ challenges and workflows of lesson adaptation, we conducted interviews with six in-service teachers and two education experts involved in both inter-regional (e.g., Shanghai-Yunnan) and intra-regional teaching support. 
Participants first used a prototype lesson plan generator as a design probe, which provoked reflections on their current practices. Our research revealed three primary obstacles that hinder effective adaptation: First, teachers \textbf{lack actionable and visible student profiles}, forcing them to rely on intuition. 
Second, existing tools is insufficient and \textbf{misaligned with their actual workflow}; 
our prototype, focused on generation from scratch, was rejected because teachers’ actual practice is often one of modification. Third, teachers face a high creative burden in manually designing culturally relevant and \textbf{contextualized learning tasks} for diverse student groups. 

Motivated by these findings, we developed \textbf{AdaPT},  an LLM-powered system that reframes lesson preparation from \textit{creation to transformation}. Instead of requiring teachers to generate entirely new materials, our system aligns with their natural preparation practices—revising and contextualizing prior plans. AdaPT includes three core features: (1) \textbf{\textit{Analyzing Student Profiles:}} The system allows teachers to input and summarize student data to generate actionable insights for adaptation. (2) \textbf{\textit{Supporting One-Click Plan Transformation:}} Instead of manual revision, the system enables teachers to automatically adapt an entire lesson plan to a new context based on the student profile analysis. (3) \textbf{\textit{Generating Differentiated Tasks:}} To reduce creative burden, the system provides multi-level and contextualized learning tasks tailored to the target students' needs and local culture. Powered by these features, AdaPT empowers teachers to efficiently and effectively adapt lessons for educational equity.

To evaluate its effectiveness, we conducted a single evaluation study with 12 participants, including 9 in-service teachers and 3 education specialists. All participants used AdaPT to complete a realistic lesson plan adaptation task. Teachers highlighted that the system significantly reduced their perceived preparation workload and was well aligned with their natural workflows. The experts, drawing on their pedagogical and research perspectives, further provided insights into the educational value of the system-generated plans, confirming their high quality and contextual appropriateness for the target school. Based on these findings, we discuss design implications for future AI-based educational technologies that support teachers’ adaptation practices and help mitigate educational inequality.

In summary, this paper makes the following contributions:

(1) We identify the key challenges teachers face in lesson plan adaptation, including data scarcity, workflow misalignment, and creative burden, derived from a formative study.

(2) We introduce AdaPT, an interactive LLM-powered system that adapts existing lesson plans to different contexts and student profiles.


(3) We conduct a mixed-methods evaluation demonstrating that AdaPT reduces teacher workload while producing high-quality, contextually relevant lesson plans, and we derive implications for designing future educational tools.


\section{Related Work}
Our study is situated in leveraging technology to narrow educational inequalities in China, with a particular focus on adapting high-quality lesson plans from high-performing contexts to meet the needs of students in lower-performing settings. We summarize prior research that 1) examines the structural and instructional inequalities that separate advantaged and disadvantaged learners in China, 2) traces the evolution of lesson plan assistance tools from structured templates to AI-powered systems, and 3) investigates the principles and challenges of human–AI collaboration in educational contexts. We identify a gap: little work has connected them to address the specific problem of adapting high-quality lesson plans from high-performing contexts to meet the needs of students in lower-performing settings. Our work addresses this gap by designing a teacher-in-the-loop, LLM-powered lesson plan transformation system aimed at promoting educational equity through contextual adaptation.

\subsection{Educational Inequality in China}
Educational inequality has been a persistent concern in China, manifesting in disparities across socioeconomic status, regions, and urban–rural divides. Prior work has provided macro-level overviews and theoretical models to explain these disparities. Li et al. \cite{li2017human} highlight the critical role of human capital in China’s economic growth, pointing out how uneven educational quality and access between urban and rural areas constrain long-term development. Hannum and Park \cite{hannum2007education} further emphasize that institutional arrangements and market reforms have reshaped the distribution of educational opportunities, often amplifying inequities. International perspectives (e.g., Ainscow \cite{ainscow2020promoting}) stress that equitable education requires system-wide approaches encompassing policy, school culture, pedagogy, and resource allocation. In this paper, we focus on educational equity rather than equality of provision. In this paper, while we do not aim to address macro-level economic or policy causes of educational inequality in China, we focus on improving educational equity at the instructional level by leveraging technology to adapt high-quality lesson plans to local contexts.

Empirical evidence consistently shows pronounced disparities between regions and between urban and rural areas. For instance, Yang \cite{yang2017research} find that urban residents enjoy significantly higher returns to education than rural counterparts, with gendered patterns of inequality. Zhang and Kanbur \cite{zhang2009spatial} document persistent spatial inequalities in education quality and healthcare access, exacerbated by fiscal decentralization and marketization. A few studies point to intertwined forms of inequity, such as the digital divide rooted in infrastructure, skills, and sociocultural factors  \cite{anderson2015digital,salemink2017rural}, and the uneven distribution of teacher quality across advantaged and disadvantaged schools \cite{goldhaber2015uneven}.

At the classroom level, inequalities in instructional quality and learning opportunities further compound these gaps. Tsang and Ding \cite{tsang2005resource} show that per-student spending varies substantially across counties, even within the same province. International assessments such as TIMSS reveal that teacher characteristics and instructional alignment can influence not only overall performance but also the equity of outcomes across socioeconomic groups  \cite{burroughs2019teacher}. Technology-based interventions, such as mobile-assisted learning \cite{khan2019mitigating}, computer-assisted learning programs \cite{mo2015computer}, and ICT-enabled collaborative teaching \cite{yang2018promoting,li2023impact}, have shown promise in mitigating some disparities.

However, the effectiveness of technology interventions is strongly shaped by local implementation capacity. For example, recent work on live-streaming-based dual-teacher classes (LSDC) in disadvantaged regions of China \cite{sun2025live} finds that while LSDC can grant students access to high-quality educational resources, its success hinges on the situated practices of local teachers, who contextualize centrally delivered lectures to fit the needs of their students. This underscores that achieving educational equity requires not only resource provision but also context-sensitive adaptation. Yet, few studies have directly addressed the challenge of systematically transforming high-quality instructional materials from high-performing contexts to meet the needs of disadvantaged learners. Our work addresses this gap by designing a teacher-in-the-loop, LLM-powered lesson plan transformation system aimed at promoting equity through contextual adaptation.

\subsection{Lesson Plan Assistance Tools}
Research on lesson plan assistance tools spans from template-driven, rule-based systems to recent large language model (LLM)-powered platforms. Traditional tools, such as iLessonPlan \cite{andre2011ilessonplan}, PLATON  \cite{strickroth2019platon}, and CIDS \cite{zain2017collaborative}, support teachers—especially novices—in structuring, visualizing, and reflecting on lesson plans through graphical interfaces, time-oriented sequences, and collaborative communities. Systems like Content Wizard \cite{chau2017content} provide concept-based recommendations by mapping course structures to instructional resources, while others integrate instructional design models to scaffold lesson preparation in multicultural or competence-based contexts \cite{calandra2007electronic,pender2022ai}. These tools improve efficiency and reflective practice, but they are typically designed for lesson creation from scratch, offering limited functionality for adapting existing plans to different learner profiles.

The emergence of LLM-based tools expands the scope of lesson planning support. Platforms such as LessonPlanner \cite{fan2024lessonplanner}  leverage generative AI to co-create pedagogy-driven plans aligned with established instructional frameworks. Some researchers have attempted to optimize AI support for specific learning scenarios—for example, Shao \cite{shao2025unlocking} used LLM to generate analogies that enhance students’ conceptual understanding. Machine learning systems such as CoFee \cite{bernius2022machine} also automate feedback on student work, further diversifying AI-assisted teaching tools. In addition, existing studies have evaluated the application of generative AI in education, finding that its acceptance in educational contexts remains relatively limited \cite{mogavi2023exploring,collie2025teachers}. 

Despite these advances, current AI-assisted lesson planning tools are still mainly designed for creating new lesson plans from scratch or for making superficial augmentations to generic templates, rather than supporting the more common teaching practice of adapting and reusing an existing high-quality plan. In real teaching practice, teachers often start from a "standard" plan—usually collaboratively developed within a school’s teaching and research group or authored by experienced teachers —and modify it to fit their own students’ needs. This gap points to the need for systems that not only emulate this adaptation process, but also align with teachers’ established workflows and preferences, thereby fostering higher acceptance in everyday practice.

\subsection{Human–AI Collaboration in Education}
The design and development of educational AI are increasingly driven by a growing corpus of surveys and frameworks. These works effectively map the conceptual and practical design spaces of multimodal interaction, including natural language~\cite{song2023gvqa}, visualizations~\cite{song2023vividgraph,lin2026camv}, and speech~\cite{chang2025rhythmta}. Kui et al. \cite{kui2022survey} synthesize a decade of visual analytics research in online education into a taxonomy of four functional categories, while Wang et al. \cite{wang2022understanding} identify key design tensions in AI-mediated student networking and propose collaborative, diverse matching strategies. Zheng et al. \cite{zheng2022ux} review conversational AI in HCI, highlighting the shift from dyadic to polyadic interaction models and the importance of boundary-awareness, trust, and ethics. Complementing these mappings, the Thinking Through AI framework \cite{tzirides2025thinking} adopts a cognitive–collaborative methodology, positioning students, teachers, and developers as co-designers. Together, these works chart the existing landscape and point to opportunities for integrating AI adaptivity with human oversight to enhance context-sensitivity, agency, and equity.

On the student side, research has examined AI-supported personalization and collaborative learning. 
Studies explore mutual theory of mind in human–AI interaction \cite{wang2021towards}, pedagogical prompting to scaffold student–AI communication \cite{xiao2025improving}, and multimodal co-design approaches for project-based learning \cite{prasad2025exploring,zheng2024charting}. 
These works show how AI can adapt content and interaction patterns to individual learners, aligning with the goal of differentiated instruction.

On the teacher side, the literature emphasizes educator control, oversight, and agency in AI integration. Systems like ReadingQuizMaker \cite{lu2023readingquizmaker} enable co-creation of assessments, while educator-centered analyses highlight potential harms, role shifts, and ethical tensions in AI–teacher partnerships \cite{harvey2025don,chng2023ai}. These insights underscore the need for designs that place teachers at the center, enabling them to guide, review, and contextualize AI outputs.

Generative AI has become increasingly prominent in recent years~\cite{xiao2024typedance,song2024gvvst}. Their applications in education—such as bias-aware content generation frameworks \cite{lim2025debiasme}, curriculum co-design systems \cite{fong2022modular}, and teacher–AI collaborative authoring tools \cite{lu2023readingquizmaker, zixin2026vizQstudio}—showcase their potential to enhance instructional design and support differentiated learning. These studies emphasize the importance of educator oversight, contextual alignment, and equity considerations in AI-assisted content creation. However, most existing tools remain task-specific and fragmented, lacking end-to-end capabilities that mirror real teaching workflows. In particular, they seldom address the pedagogical adaptation required to migrate high-quality instructional resources from high-performing contexts to lower-performing ones—a gap closely tied to educational inequities. Bridging this gap by integrating teachers’ pedagogical expertise with AI’s generative flexibility for lesson plan transformation presents an underexplored opportunity to advance educational equity in real-world classrooms.

\section{Formative Study}
To ground the design of AdaPT in real-world teaching contexts, we conducted a formative study with teachers and education experts. 
The study aimed to uncover how lesson plans are currently adapted across diverse student groups, identify the challenges teachers face in this process, 
and elicit requirements for an AI-assisted system to better support lesson plan transformation.

\subsection{Participants}
We conducted semi-structured interviews with eight participants, including two education experts (E1–E2) and six in-service teachers (T1–T6). 
Their teaching experience ranged from 5 to over 20 years, covering multiple subjects such as Mathematics, Chemistry, and Information Technology. Participants included an expert in educational informatization research (E1), a vice principal with rural teaching experience (E2) and teachers from both high-performing and lower-performing schools. 
All participants had experience teaching multiple student groups with diverse learning profiles and had engaged in adapting lesson plans to accommodate the needs of different groups. 
Prior to the interviews, the six teachers were invited to use a structured lesson plan generation system, after which we collected their feedback and subsequently conducted the semi-structured interviews.
The detailed background of the participants is summarized in Table~\ref{tab:participants}.

\begin{table}[h]
  \caption{Background of Formative Study participants.}
  \label{tab:participants}
  \begin{tabular}{cccccl}
    \toprule
    ID & Years & Intra-school & Intra-regional & Cross-regional & Notes \\
    \midrule
    E1 & 20--25 & \checkmark & \checkmark &  & Education expert \\
    E2 & 20--25 & \checkmark & \checkmark & \checkmark & Vice principal \\
    T1 & 20--25 & \checkmark & \checkmark &  & LMS use leader \\
    T2 & 25--30 & \checkmark &  &  &  \\
    T3 & 25--30 & \checkmark & \checkmark &  &  \\
    T4 & 5--10  & \checkmark & \checkmark &  &  \\
    T5 & 10--15 & \checkmark & \checkmark &  &  \\
    T6 & 10--15 & \checkmark &  & \checkmark &  \\
    \bottomrule
  \end{tabular}
\end{table}

\subsection{Procedure}

\begin{figure}[h]
  \centering
  \includegraphics[width=\linewidth]{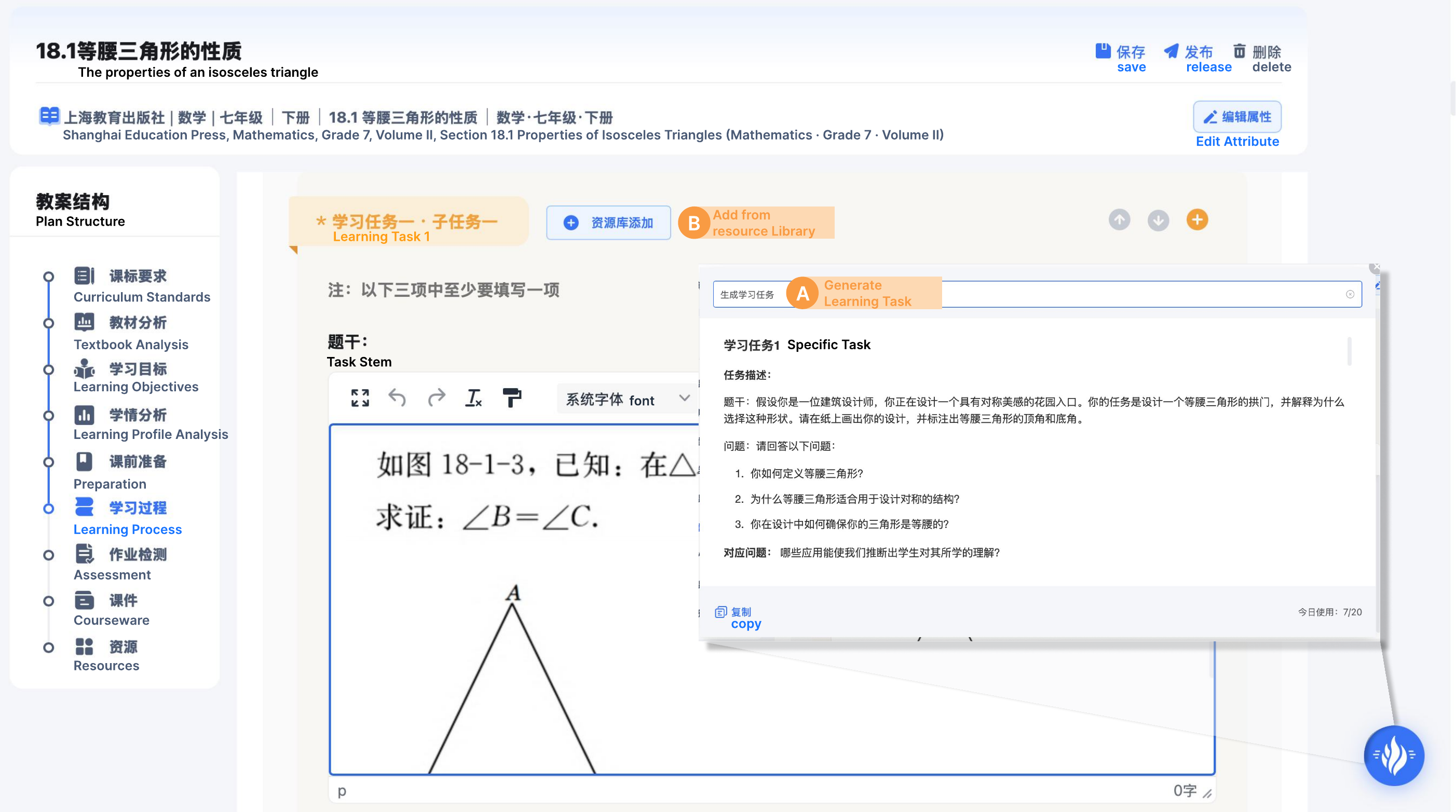}
  \caption{The structured lesson plan authoring system used in Stage 1 of the formative study.}
  \Description{}
  \label{fig2}
\end{figure}

The study followed a two-stage procedure. 
In the first stage, we conducted a system use experiment in which every participants were asked to use a structured lesson plan authoring tool. 
The system was organized into five main modules: \textit{learning objectives}, \textit{student profile analysis}, \textit{textbook analysis}, and \textit{pre-class}, \textit{in-class}, and \textit{post-class learning tasks}. 
The learning objectives had been pre-populated in the system by decomposing China’s national curriculum standards into granular objectives. 
The student profile analysis module was completed subjectively by teachers. 
For the textbook analysis and learning task modules, the learning tasks from the textbooks used in the target region had been pre-entered into the system by the research team (\autoref{fig2}-B), but teachers were required to produce corresponding lesson plans and teaching scripts for each task.

To assist this process, we integrated an AI agent into the learning task module, offering default functions such as ``generate task analysis'', ``generate teaching script'', and ``evaluate learning task'' (\autoref{fig2}-A). 
All six teachers were assigned to adapt lesson plans originally designed for their own schools to be used in another local school with significantly weaker student profiles. 
The goal was to redesign the lesson plans in the system to suit the target school’s context.

The second stage involved semi-structured interviews with all participants. 
Participants reflected on their experiences using the system, described their strategies for adapting lesson plans to different student groups, and shared their expectations for future system features.

Interviews focused on four themes:
\begin{enumerate}
    \item \textit{Experience with AI-generated lesson plans} – Participants evaluated the quality of AI-generated content across different subjects, identifying strengths in structured domains such as algebra and weaknesses in areas requiring graphical representation. They also discussed their workflow when integrating AI outputs into teaching materials, including post-editing and using additional AI tools for refinement.

    \item \textit{Lesson plan adaptation based on student profiles} – Participants described how they modify lesson plans according to students’ grade level, academic performance, cognitive ability, and engagement level. Adjustments typically involved changing task difficulty, altering the ratio of foundational to extended activities, and modifying the teaching format.

    \item \textit{Cross-school or cross-class migration} – Participants reflected on experiences transferring lesson plans between schools or classes with varying student conditions. All agreed on the necessity of adjustments when student ability levels diverged significantly, with STEM subjects more amenable to difficulty scaling than humanities subjects. They also noted that information literacy disparities among teachers influence how effectively technology-supported plans can be adopted.

    \item \textit{System feature expectations} – Participants suggested features to lower adoption barriers, such as automatically generating multi-level versions of the same lesson, standardizing the lesson plan format for quick comprehension, and enabling direct ``plug-and-play'' use to counter teacher reluctance. They also discussed the role of institutional support, stressing that technology must be complemented by policy incentives and quality assessment mechanisms.

\end{enumerate}

\subsection{Challenges Identified}
The challenges identified in this study emerged from the two-stage procedure: a system use experiment with a structured lesson plan authoring system and subsequent semi-structured interviews with all participants. Through participants’ reflections on lesson plan adaptation across diverse student groups and school contexts, we distilled five core challenges that informed our design requirements.

\subsubsection*{\textbf{C1: Lack of Systematic Student Profile Data.}}
Teachers often lack systematic and reliable data to inform lesson plan adaptation. As T5 explained, ``\textit{Some subjects such as chemistry were not introduced until the 8\textsuperscript{th} grade for students.   I have no past exam scores or performance records to look at.}'' Lesson planning relies heavily on personal experience and analogy from other subjects.  In contrast, for subjects with a longer history, teachers still relied on limited metrics. T3 reported, ``\textit{As a teacher of a core subject or a homeroom teacher, we usually teach students from enrollment to graduation. However, we often only adjust lesson plans based on existing grades, and cannot comprehensively compare the differences in student situation.}'' All teachers acknowledged that current practices depend on individual observation, teaching experience, and feedback, with no unified standards. This problem is amplified in cross-school migration scenarios. As E2 noted, ``\textit{Local teachers’ ICT skills vary greatly, and there is no solution to quickly compare the learning profiles between schools,}'' making it difficult for incoming teachers to adapt efficiently.

\subsubsection*{\textbf{C2: Difficulty in Translating Student Differences into Targeted Changes}}
When transferring lesson plans across schools or classes, teachers must clearly understand student differences to make targeted changes. T1 recounted her experience moving from a high-performing to a lower-performing school: 
``\textit{When I first moved, I was confident in my lesson plans, but it was a disaster. I’d give them homework that I thought was standard, and most of the class would turn it in blank. They weren't being lazy; the material was just too difficult for them. It took me weeks to realize I couldn’t just tweak things; I had to fundamentally rethink and simplify everything.}''  This sentiment was echoed by other in-service teachers,  who agreed that while significant differences in learning profiles demand adjustments, such differences are rarely visualized in a way that makes them actionable. T2 commented, ``\textit{We all know the classes are different. But how different? Is it a 10\% difference in comprehension, or 50\%? The data we have doesn’t tell us what to change.}'' Therefore, for frontline teachers, even though there is data, it is difficult to turn it into practical and actionable changes.

\subsubsection*{\textbf{C3: Low Adoption of Structured Lesson Plans}}
In many under-resourced areas, teachers rarely use standardized, structured lesson plans, leading to inconsistent quality and making cross-context adaptation highly challenging. E2 observed that in Yunnan, lesson plans are often written without a unified structure, and when no prescribed format is followed, the overall quality tends to be low. This lack of structured representation not only hampers teaching logic but also increases the burden of adapting plans for new contexts. T3 suggested that lesson plans should first be converted into a unified, structured ``\textit{platform format}'' to enable easier customization. E2 and T5 further emphasized that structured formats can support clearer pedagogical flow and make it easier for teachers with limited ICT skills to engage. The need for ``\textit{one-click usable}'' structured lesson plans is particularly pressing in lower-performing and rural schools, where teacher engagement with digital tools is limited. T1 also highlighted the value of auto-generated, lower-difficulty versions to reduce extensive manual revision.

\subsubsection*{\textbf{C4: Difficulty in Crafting Contextualized Learning Tasks.}}
Teachers face considerable challenges when designing learning tasks that address not only the diverse abilities of students but also their cultural and situational contexts. While tasks for advanced students often emphasize creativity and challenge, those for students with weaker foundations focus on consolidation and incremental learning. E1 shared, ``\textit{When designing learning tasks, embedding cultural elements familiar to students usually results in higher engagement. For instance, incorporating local traditions into lessons makes the classroom interaction noticeably better.}'' However, conducting research and designing such culturally relevant tasks demands extra effort from teachers.  T4 further added, ``\textit{We have to spend additional time learning about students' interests and integrating those elements into tasks. This becomes particularly difficult when managing diverse student groups across schools.}'' In one case, E2 redesigned tasks around the locally popular animated series Kung Fu Panda to increase engagement and emotional connection. Teachers also called for greater diversity in task formats, such as project-based learning, group collaboration, or ``\textit{peer support}'' structures within groups. However, they currently lack ready-made templates for multi-level, contextualized tasks, leading to time-consuming manual splitting and rewriting.

\subsubsection*{\textbf{C5: Fragmented and Non-iterative Refinement Process.}}
In practice, AI-generated lesson plans serve only as first drafts requiring multiple rounds of revision. All teachers reported post-editing AI outputs; T3 would ``\textit{generate with one AI, then rewrite from another perspective with a different AI,}'' while T5 used in-system prompts (e.g., the ``\textit{little flame}'' feature) to broaden lesson ideas.  Similarly, T2 remarked, ``\textit{I wish I could complete all design and editing tasks within one system, rather than jumping between multiple platforms. It’s inefficient and frustrating.}'' E2 proposed a potential solution: ``\textit{If the system could support multi-round iterations, such as generating a draft, optimizing it based on teacher feedback, and producing a refined version, it would greatly improve our experience.}'' In summary, teachers require a system that eliminates fragmented workflows and supports continuous, in-system refinement and collaboration, ensuring a seamless and efficient lesson plan design process.

\begin{figure}[h]
  \centering
  \includegraphics[width=\linewidth]{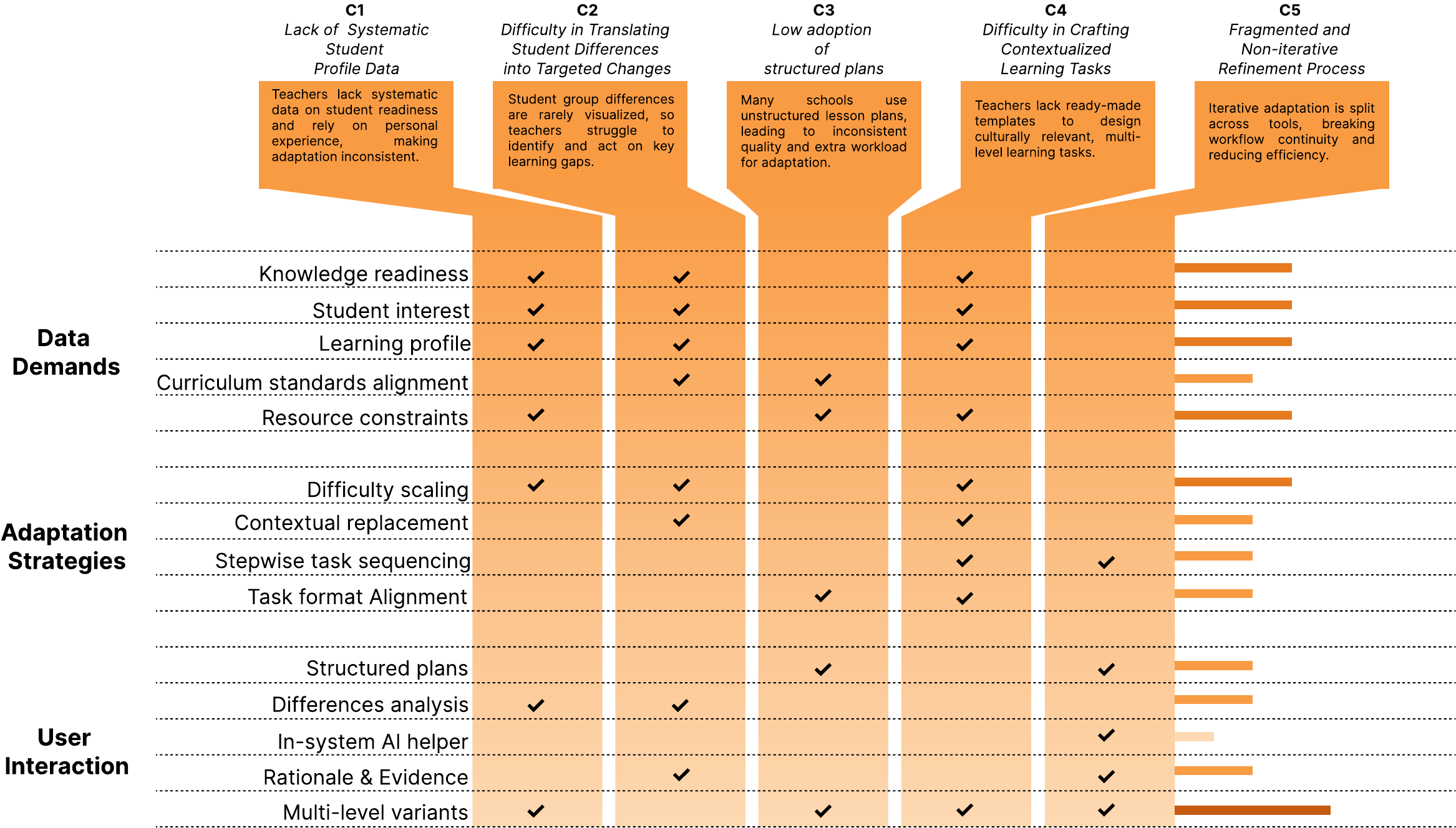}
  \caption{Compiled results of the formative study. The five identified challenges (C1–C5) are mapped against three categories of system support: Data Demands, Adaptation Strategies, and User Interaction.}
  \Description{}
\end{figure}

\subsection{Design Requirements}
Based on the insights gained from the formative study, we identified several key requirements for the development of an AI-assisted lesson plan transformation tool that facilitates a human-AI collaborative process in adapting and redesigning lesson plans for different student learning profiles.

\subsubsection*{\textbf{DR1: Enable Flexible and Structured Entry of Student Data}}
To address the lack of systematic student profile data (\textbf{C1}), the system must first focus on capturing and structuring teachers' knowledge of their students. This requires providing flexible input mechanisms that accommodate teachers' existing habits, such as allowing them to describe their class in natural language or manually enter key observations. The system's primary role here is to parse these varied inputs and synthesize them into a holistic and structured student profile, identifying key attributes like prerequisite knowledge gaps, learning pace, and specific interests. Such flexibility would allow for more precise and actionable analyses, ensuring that adaptations are sufficiently tailored to diverse teaching environments.

\subsubsection*{\textbf{DR2: Bridge the Gap from Analysis to Action with Automated Suggestions.}}
Building upon the structured profile created, the system should help teachers clearly see differences in students’ knowledge readiness, learning styles, student interests and etc. Beyond making such differences visible, the system should also provide actionable modification suggestions (\textbf{C2}) and directly auto-generate adapted lesson plans. This end-to-end support would give teachers concrete starting points for targeted adjustments while significantly reducing the manual effort required.

\subsubsection*{\textbf{DR3: Automate the Conversion to a Structured and Editable Format.}}
To overcome the low adoption of structured plans in under-resourced contexts (\textbf{C3}), the system need to automate the conversion of existing, often unstructured, lesson plans into a standardized and easily editable format. This includes the ability to import plans from various sources (e.g., text documents) and intelligently parse them into meaningful pedagogical sections like \textit{learning objectives}, \textit{knowledge points}, and \textit{learning tasks}. By handling the structural heavy lifting, the system will lower the barrier for teachers with varying ICT skills, promote a consistent quality standard, and make the subsequent adaptation process more efficient.

\subsubsection*{\textbf{DR4: Generate Multi-Level and Contextually-Aware Learning Tasks.}}
To alleviate the high creative burden of designing contextualized learning tasks (\textbf{C4}), the system should support not only differentiated difficulty levels but also cultural and situational relevance. This would help teachers design tasks that resonate with students’ interests and local contexts, while offering multi-level templates to reduce preparation time.

\subsubsection*{\textbf{DR5: Support In-System Iteration and Refinement.}}
To address the fragmented and inefficient refinement workflow (\textbf{C5}), the system should support a in-system iterative process. It should move beyond a one-shot generation model to enable a  human-AI co-creation loop. The system should then revise the plan directly within the same interface, eliminating the need to export to external tools.

\section{System}
In this section, we first outline the overview of AdaPT through a representative user example. We then detail the internal representation of lesson plans, which establishes the schema for curriculum standards, learning objectives, learner profile analysis, and learning tasks. 
Based on this representation, we further present LLM-based methods for lesson plan transformation and the visual interface.


This section begins with a walkthrough of a typical user experience with AdaPT, followed by a detailed explanation of each system component and how it addresses the identified design requirements.

\subsection{Example User Walkthrough}

Jason is a high school information technology teacher from a metropolitan city. As part of a rural teaching support program, he brings with him a set of standard lesson plans originally designed for urban students. Upon arriving in the rural school, Jason quickly realizes that the students’ learning profiles differ significantly: many have limited prior exposure to computers, fewer opportunities for collaborative learning, and their interests are shaped by agricultural and local community contexts. Although local teachers provide him with some descriptions of student readiness, these descriptions are broad and lack sufficient detail for him to confidently adapt his lesson plans.

\begin{figure}[h]
  \centering
  \includegraphics[width=\linewidth]{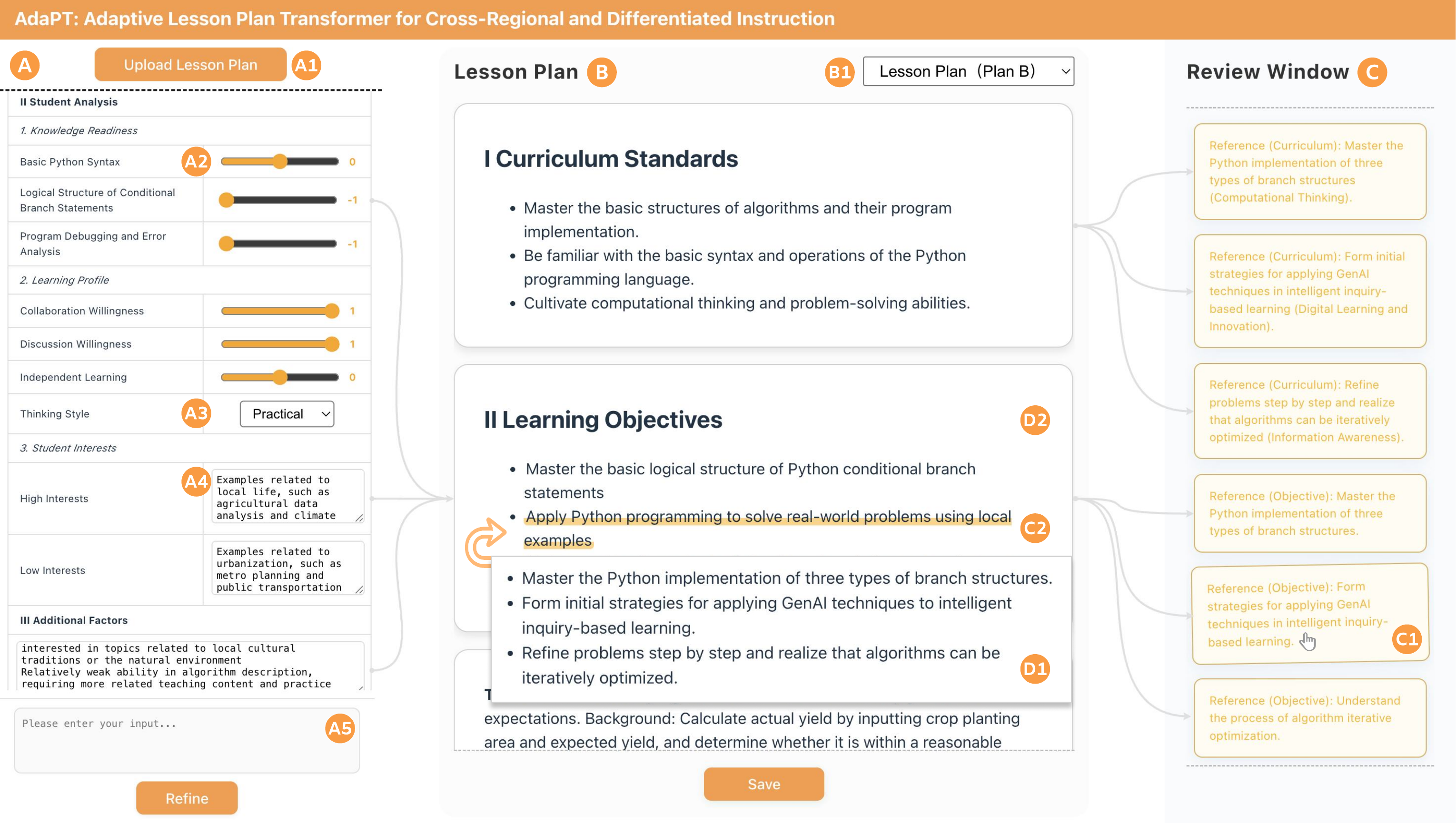}
  \caption{
  The user interface of AdaPT. The interface is composed of three coordinated views. (A) The \textit{Profile Panel} for uploading lesson plans and configuring student learning profiles through structured inputs, (B) The \textit{Lesson Plan Workspace} for editing, adapting, and comparing different versions of instructional content, and (C) The \textit{Review Window} for inspecting AI-generated justifications and references to previous lesson plans.
  }
  \Description{}
  \label{fig4}
\end{figure}

To address this, Jason turns to AdaPT. He uploads his original lesson plan into the system (\autoref{fig4}-A1) and provides a short prompt describing the rural teaching context (\autoref{fig4}-A5). By clicking the Summarize button, the system automatically generates a structured student profile analysis, combining general knowledge readiness with context-specific factors, and presents it in a tabular view (\autoref{fig4}-A). Jason reviews this AI-generated profile, edits entries where needed, and adds his own observations.

With the completed profile in place, Jason clicks the Transform button. The system converts the uploaded lesson plan into a new, adapted version that reflects the rural students’ learning needs (\autoref{fig4}-B). For instance, examples involving urban transportation are replaced with local agricultural scenarios.

Jason can then directly compare the transformed plan against the original (\autoref{fig4}-B1), with highlighted differences (\autoref{fig4}-C2) and AI-generated explanations for each modification (\autoref{fig4}-C). If he finds any changes unsatisfactory, he can either manually edit the plan or enter additional prompts to guide the AI in refining the content. Through this iterative process, Jason efficiently develops a lesson plan that is pedagogically consistent yet tailored to the new teaching environment, ensuring both relevance and instructional quality.

\subsection{Lesson Plan Representation}
To enable systematic transformation of lesson plans, we designed a structured representation that decomposes a lesson plan into four key components: \textit{curriculum standards}, \textit{Learning Objectives}, \textit{Learning Profile Analysis}, and \textit{Learning Tasks}. This structure builds upon both theoretical foundations from curriculum studies and practical insights gathered through expert interviews.

Our design draws inspiration from the \textit{Encyclopedia of Curriculum Studies} \cite{kridel2012instructional}, which highlights four common features across instructional design models: (1) they involve some level of analysis, (2) they address organization, (3) they consider delivery of instruction, and (4) they incorporate means for evaluation. Guided by this framework, we developed a schema that allows lesson plans to be parsed, adapted, and restructured while preserving their pedagogical integrity.

First, the \textit{Curriculum Standards} represent nationally mandated requirements published by the Ministry of Education, defining the learning outcomes that every student must achieve. 

Second, the \textit{Learning Objectives} specify the intended instructional goals based on standards for a particular lesson, reflecting the teacher’s concrete expectations for student progress within the given class session.

Third, the \textit{Learning Profile Analysis} in our system operationalizes the principles of \textit{Differentiating Instruction in Response to Student Readiness, Interest, and Learning Profile in Academically Diverse Classrooms: A Review of Literature} \cite{tomlinson2003differentiating}. While the literature frames this as ``differentiated instruction'', we adapt the terminology to \textit{Learning Profile Analysis} to better align with teachers’ everyday practice. Within this framework, the specific sub-dimension of \textit{learning profile} directly corresponds to Tomlinson’s formulation.

For readiness, we draw on Vygotsky’s theory of the \textit{zone of proximal development} \cite{vygotsky1978mind,vygotsky1986thought}, which highlights the teacher’s responsibility to guide learners into their developmental zone by supporting tasks slightly above their independent performance level. To enable this, our system explicitly analyzes students’ \textit{knowledge readiness}, identifying prerequisite gaps that inform appropriate scaffolding and progression.

We further incorporate indicators of \textit{student interest}, recognizing its central role in sustaining motivation and engagement \cite{amabile2018creativity,csikszentmihalyi1997talented}. In addition, the \textit{learning profile} dimension captures preferences and tendencies relevant to instruction, including collaboration, discussion, independent learning, and preferred modes of thinking (analytical, practical, or creative). To support teachers’ reflective practice, we also include an \textit{Additional Factors} field for contextual variables (e.g., rural vs. urban settings, access to digital resources) that may shape instructional choices. This multidimensional representation echoes findings from our formative study, in which teachers emphasized considering not only prerequisite knowledge but also social and contextual constraints when adapting lesson plans.

Fourth, \textit{Learning Tasks} are refined into structured units consisting of the \textit{task} itself, its expected \textit{answer}, a \textit{detailed plan}, the suggested \textit{implementation method}, and the underlying \textit{design intent}. This granular structure serves two purposes: first, it allows teachers to directly reuse tasks in their classrooms; second, it facilitates reflection on whether each task aligns with the mandated curriculum standards and stated learning objectives. In doing so, tasks become the actionable bridge linking standards, objectives, and student profiles, enabling systematic yet teacher-centered adaptation.

\subsection{User Interface Design}
In the following section, we introduce core views and interaction flow of the AdaPT user interface.

\subsubsection{Single Views}
The AdaPT user interface is architecturally centered on three coordinated views: the \textit{Profile Panel} (\autoref{fig4}-A), the \textit{Lesson Plan Workspace} (\autoref{fig4}-B), and the \textit{Review Window} (\autoref{fig4}-C), with additional interface elements such as import and export buttons supporting extended functionality. 
This layout supports teachers in capturing learning profile context, transforming and editing instructional content block by block, and inspecting AI generated evidence alongside corresponding plan elements. 

The left side of the screen is the \textit{Profile Panel}. At the top-left corner of the system is a button for uploading original lesson plans (\autoref{fig4}-A1).
The Profile Table under the button supports flexible encoding of student characteristics through structured inputs such as Knowledge Readiness sliders (e.g., setting Basic Python Syntax to middle readiness, \autoref{fig4}-A2), Learning Preferences dropdowns (e.g., selecting a thinking style, \autoref{fig4}-A3), and open text fields for contextual factors (e.g., student interests in agriculture, \autoref{fig4}-A4). 
These inputs are instantly serialized into a structured profile, combining subjective and objective data into actionable analyses (\textbf{DR1}). 
The central \textit{Lesson Plan Workspace} renders the lesson as semantically grouped blocks including Curriculum Standards, Learning Objectives and Learning Tasks, each editable in place with Markdown support and targeted regeneration features. 
For instance, a teacher may modify a programming example in one block while triggering AI assisted revision only for that section, maintaining coherence without disrupting adjacent content(\textbf{DR5}). 
The \textit{Review Window} on the right exposes AI reasoning through visually anchored references of previous lesson plans and reasons why the parts are modified, for example showing a justification such as lowered difficulty due to weak debugging skills linked to a modified exercise, enabling transparent validation and one click incorporation of suggestions. 
Together, these views operationalize requirements from users by making learning gaps visible, automating structural conversion, supporting contextualized task design, and retaining teacher agency throughout.

\subsubsection{Interaction Flow}
AdaPT orchestrates a seamless human AI collaboration workflow through an integrated interaction sequence that begins with profile summarization and progresses through transformation, inspection, and iterative refinement. 
Teachers first click the upload button (\autoref{fig4}-A1), sending the raw lesson plan to AdaPT. 
Due to variations in student readiness across regions, the original lesson plan cannot be directly applied in the current teaching context. 
Therefore, the teacher should enter a description of the students’ current learning situation into the text input box at the bottom left as a natural language prompt (\autoref{fig4}-A5), and then trigger the Summarize function to generate a structured analysis. 
With the assistance of LLM, the system can derive student profiles according to the description of the previous lesson plan. 
Then the structural student profiles are presented in the left panel with the interactive components which can be used to modify the generated profiles, combining sliders, dropdowns, and free text prompts. 
Upon clicking the Transform button, the system automatically converts the original lesson plan into a standardized editable structure aligned with the profile. 
For example, it can convert a generic programming lesson into agriculture themed examples for students with related interests. 
When user hovers on the items in lesson plans or references (\autoref{fig4}-C1), the related content in the central workspace will be highlighted (\autoref{fig4}-C2), showing differences between original and adapted versions (\autoref{fig4}-D1\&D2).
With the comparison between different versions of lesson plans, teachers can inspect AI proposed changes block by block while reviewing evidence in the right pane. 
They might see a suggested replacement activity justified by both knowledge gaps and cultural relevance. 
Teachers can accept, modify, or regenerate specific sections using contextual prompts, such as requesting to simplify an explanation for beginners. 
They can also switch between alternative plan variants with version control dropdowns (\autoref{fig4}-B1), such as Plan A versus Plan B, to compare adaptation strategies, and repeatedly refine the draft without leaving the system. 
The flow concludes with final saving or exporting, ensuring all revisions remain traceable and teacher controlled. 
This end to end process effectively improves the workflow by automating structural adaptation, reducing manual effort, supporting contextualized customization, and enabling multi step refinement within a unified environment.

\subsection{LLM-based Lesson Plan Transformation}
The backend of AdaPT is designed as a multi-agent system (\autoref{fig5}), where each agent encapsulates distinct reasoning responsibilities aligned with the design requirements. This agent-oriented architecture not only decomposes the complex workflow of lesson plan transformation into manageable stages, but also ensures that outputs remain interpretable, editable, and pedagogically grounded. The two primary agents are the \textit{LessonPlan Agent} and the \textit{LearningProfile Analysis Agent}.

\subsubsection{LessonPlan Agent}
The \textit{LessonPlan Agent} is responsible for processing lesson plan workflows, covering both the structuring of original lesson plans and their subsequent transformation. 
In the structuring stage (\autoref{fig5}-A), the agent parses an uploaded lesson plan into schema-conformant components (\textbf{DR3}): 
\begin{itemize}  
    \item \textbf{Curriculum Standards}: retrieved and validated through a RAG process. The agent queries a tree-structured knowledge graph of national curriculum standards, embedding knowledge points for semantic alignment.  
    \item \textbf{Learning Objectives}: extracted from the original plan as teacher-specified instructional goals.  
    \item \textbf{Learning Profile}: reformatted into a structured schema where knowledge readiness remains initially unknown but is subsequently enriched with knowledge graph nodes related to curriculum standards. The agent verifies and supplements these entries to ensure completeness.  
    \item \textbf{Learning Tasks}: normalized into atomic units for later comparison and refinement.  
\end{itemize}

In the transformation stage (\autoref{fig5}-C), the \textit{LessonPlan Agent} adjusts the parsed plan according to the inferred student profile. The agent first recalibrates \textit{Learning Objectives} in light of profile differences, then formulates a high-level \textit{Transformation Overview Idea} that specifies adaptation strategies (\textbf{DR2}). Finally, it iteratively adjusts each \textit{Learning Task} according to targeted suggestions, including difficulty calibration, contextual replacement, and task scaffolding (\textbf{DR4, DR5}). This design makes learning gaps explicit, generates adapted drafts automatically, and supports iterative refinements throughout the workflow.

\subsubsection{LearningProfile Analysis Agent}
\begin{figure}[h]
  \centering
  \includegraphics[width=\linewidth]{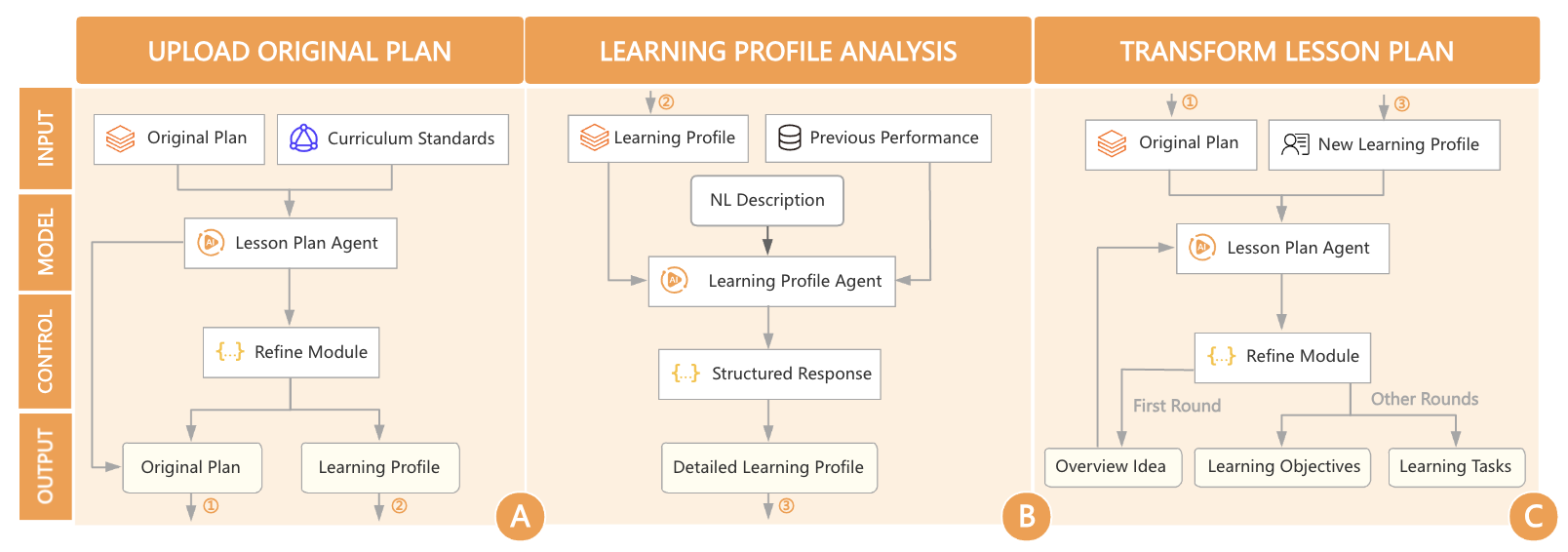}
  \caption{System backend workflow of AdaPT: (A) In the Upload Original Plan component, teachers provide the original lesson plan, which is parsed by the Lesson Plan Agent with support from a curriculum standards knowledge graph. The output is structured into base information, curriculum standards, learning objectives, learning profiles, and learning tasks. (B) In the Learning Profile Analysis component, teacher-provided natural language descriptions and students’ previous performance are processed by the Learning Profile Agent. The agent produces structured outputs including knowledge readiness, student interests, and additional contextual factors such as region, culture, and infrastructure. (C) In the Transform Lesson Plan component, the Lesson Plan Agent integrates structured objectives and profile data to generate transformation overview ideas, which are iteratively refined into adapted objectives and learning tasks, ensuring contextualized and pedagogically aligned lesson plans.}
  \Description{}
  \label{fig5}
\end{figure}

Complementing the \textit{LessonPlan Agent}, the \textit{LearningProfile Analysis Agent} specializes in profile reasoning and enrichment (\autoref{fig5}-B). Given teacher-provided natural language descriptions of the teaching context (\textbf{DR1}), the agent generates a structured profile that includes (\textbf{DR3}):
\begin{itemize}
    \item \textbf{Base Information}: class metadata (e.g. subject, grade and number of Students) derived from teacher input.  
    \item \textbf{Knowledge Readiness}: inferred from teacher descriptions or, when available, automatically retrieved from the school’s LMS through tool-calling interfaces.  
    \item \textbf{Learning Profile}: dimensions such as collaboration, discussion, independent learning, and preferred thinking styles (analytical, practical, creative).  
    \item \textbf{Student Interests}: expressed at high/low granularity.  
    \item \textbf{Additional Factors}: contextual aspects such as region, culture, economic conditions, and hardware availability.  
\end{itemize}

This reasoning process explicitly considers multiple dimensions—regional disparities, cultural relevance, economic background, and infrastructural constraints—to produce a nuanced and context-aware student profile. Outputs are returned as schema-validated JSON, ensuring consistency and enabling comparison across versions. By supporting flexible natural language input (\textbf{DR1}) and producing actionable profiles that directly inform task adaptation (\textbf{DR2, DR4}), the agent enables teachers to integrate both subjective observations and objective evidence into the transformation pipeline.

\subsubsection{Human-AI Collaboration and Traceability}
Throughout the workflow, agents generate outputs that are both structured and explainable. Each transformed component is linked back to its source span in the original lesson plan, accompanied by \textit{References} (aligned standards or evidence snippets) and \textit{Reasons} (localized rationales). This traceability not only empowers teachers to selectively adopt AI-suggested changes, but also reinforces human authority in the decision loop, thereby addressing DR5. Together, the LessonPlan Agent and the LearningProfile Analysis Agent enable an end-to-end yet teacher-centered transformation process that balances automation with pedagogical control.

\subsection{Implementation}
AdaPT is implemented as a full-stack web application. 
The frontend is built with Vue.js (Vue 3), providing an intuitive and responsive user interface. The backend is implemented in TypeScript using NestJS, which hosts and orchestrates the LLM agents and transformation services. 
The frontend and backend communicate via RESTful APIs with schema-validated JSON payloads, enabling smooth integration between the user interface and the LLM-powered analysis and transformation modules.

\section{Evaluation}
We conducted an evaluation study to examine the effectiveness of the human-AI collaborative workflow proposed in AdaPT, as well as the performance of its core components. Our primary goal was to assess to what extent the workflow, with the assistance of Large Language Models (LLMs), can improve teachers’ efficiency and reliability when adapting lesson plans. To guide our study, we formulated the following research questions:
\begin{enumerate}
\item \textit{RQ1}: Can the system effectively transform original lesson plans into structured lesson plans?

\item \textit{RQ2}: Can the system assist teachers in accurately analyzing learning profile differences, particularly in synthesizing macro-level disparities into concrete learning needs?

\item \textit{RQ3}: How efficient and effective is the system in transforming learning objectives and tasks?

\item \textit{RQ4}: How do the system-generated feedback and explanations (traces and reasons) support teachers in iterating and refining lesson plans?

\item \textit{RQ5}: How do teachers perceive the overall usability and utility of the system throughout the adaptation process and in generating the final lesson plans?

\end{enumerate}

\subsection{Procedure of User Study}
Participants were recruited via email invitations from local K-12 schools and research institutes. The study protocol received Institutional Review Board (IRB) approval. After obtaining informed consent, we collected demographic information anonymously, including teaching experience, professional role, years of practice, and prior attitudes toward using LLMs (e.g., ChatGPT) for lesson plan adaptation.

Each study session lasted approximately 60 minutes and consisted of three stages: (1) a 40-minute system exploration and lesson plan adaptation experiment, (2) a 10-minute questionnaire evaluating perceptions of each workflow step, and (3) a 10-minute post-study interview to elicit qualitative feedback. Before beginning the experiment, participants were provided with a brief introduction and system demonstration, covering how to upload lesson plans, input learning profiles, generate adapted plans, and apply manual adjustments.

\begin{figure}[h]
  \centering
  \includegraphics[width=0.7\linewidth]{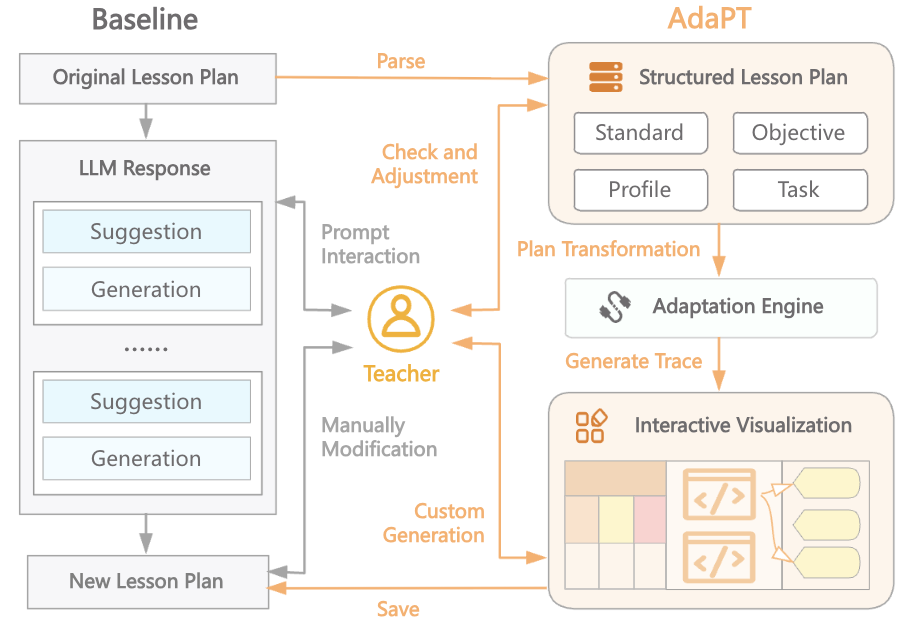}
  \caption{Overview of the user study procedure: Comparison between the baseline workflow of lesson plan modification using a single LLM and the AdaPT system workflow. }
  \Description{}
  \label{fig6}
\end{figure}

The lesson plan materials were drawn from real-world standardized K-12 IT lesson plans co-developed by a school-level preparation team. We obtained consent from both the school and the institutes, and the data collection was IRB-approved. 
Participants were asked to adapt these lesson plans to the context of a weaker class they had experience teaching (e.g., a rural support class in another region, a lower-performing class within the same district, or a weaker parallel class within their school). 
In addition, as a baseline condition, we asked teachers to independently adapt the same lesson plan using GPT, allowing us to compare the performance of AdaPT against general-purpose LLM use (\autoref{fig6}). 

To analyze teacher-AI interactions, we logged participants’ natural language descriptions of learning profiles, their manual modifications of these profiles, and their edits to AI-generated content.

\subsection{Participants}
We recruited 12 participants (P1–P12, 7 male and 5 female). Among them, three were domain experts (teaching researchers or members of the national curriculum standards committee members), and three had prior cross-regional teaching experience (e.g., from Shanghai to Yunnan). On average, participants had 12.83 years of K-12 teaching experience. All participants had previously experimented with using LLMs to generate or modify lesson plans. Each participant spent about one hour in the study and received \$100 compensation.

\begin{table}[h]
  \caption{Background of User Study participants (P1--P12).}
  \label{tab:participants-study}
  \begin{tabular}{cccccc}
    \toprule
    ID & Teaching Years & Cross-regional Exp & Notes \\
    \midrule
    P1   & 13 &  &  \\
    P2   & 1  &  &  \\
    P3   & 7  &  &  \\
    P4     & 23 & \checkmark &  \\
    P5   & 12 & \checkmark &  \\
    P6   & 10 &  &  \\
    P7   & 20  &  & Curriculum Standards committee member \\
    P8     & 5 &  &  \\
    P9     & 19 &  & Curriculum Standards committee member \\
    P10  & 10 &  &  \\
    P11  & 14 & \checkmark &  \\
    P12  & 20 &  & IT Teaching Researcher \\
    \bottomrule
  \end{tabular}
\end{table}

\subsection{User Study Results}
In this section, we present findings from the user study, structured along the five research questions. We combine quantitative results from the 7-point Likert questionnaires with qualitative insights from participant interviews. Overall, teachers evaluated the system positively across all dimensions, with mean ratings consistently above the neutral baseline of 4.
For RQ1, RQ2, RQ3, and RQ5, we set a score of 4 as the baseline representing teachers’ experience with lesson plan adaptation assisted by ChatGPT; scores higher than 4 indicate that AdaPT outperforms the baseline.
For RQ4, since ChatGPT does not inherently provide modification traces or reasons, the 7-point Likert scale was instead used to evaluate the explanatory value that teachers perceived from AdaPT’s traces and reasons, with higher scores reflecting greater clarity and support for understanding.

\subsubsection{RQ1: Structured Completeness of Transformed Lesson Plans}

Participants rated the completeness and consistency of the transformed lesson plans highly. The system was considered effective in ensuring that all core components (learning objectives, tasks, resources, implement methods, differentiation strategies, and adaptation rationales) were clearly represented. Quantitatively, teachers gave an average rating of 6.83 for completeness (Q5), 6.58 for logical coherence (Q6), and 6.33 for formatting consistency (Q7), all significantly above the baseline of 4.

Interview data further corroborated these findings. As P2 explained, compared with her usual lesson plan writing, the system’s structured transformation was \textit{``more concise, while still preserving the key information,''} and the segmentation of learning tasks was \textit{``remarkably precise.''} This reflection captures a broader pattern across the interviews: nearly all frontline teachers highlighted that the system’s structuring process distilled essential elements without redundancy, which they regarded as a significant improvement over their own manual lesson plan practices.

\subsubsection{RQ2: Alignment with Learning Profiles}
Teachers also recognized the system’s ability to adapt lesson plans to different student profiles. Ratings were consistently positive: 5.83 for alignment of objectives with student level (Q8), 5.67 for task difficulty (Q9), 6.08 for adjustments across strong–weak or urban–rural contexts (Q10), and 6.17 for responsiveness to profile differences (Q11).

Interview accounts further substantiated these results. As P11, who had prior experience teaching across regions, reflected, the system’s cross-regional profiling was ``\textit{consistent with expectations},'' noting that when she entered only the regional context, ``\textit{the system was indeed able to infer and summarize potential directions for adjustment based on contextual differences}.'' Similarly, P10 remarked that the analysis of student profiles ``\textit{gave me more ideas and reminders about how to adjust the lesson plan}.''

When examining the transformed lesson plans, most teachers affirmed that the conversion of learning objectives and tasks aligned well with their expectations. By contrast, teachers described that directly prompting GPT often fell short in differentiating levels of difficulty. P3 explained that teachers typically rely on ``\textit{behavioral verbs, such as know, understand, and master, to distinguish among objectives},'' but such nuance was often missing from GPT’s raw outputs. P6 similarly observed that GPT ``\textit{struggled to capture the varying difficulty of learning tasks},'' whereas our transformation methods more clearly separated tasks by difficulty. Teachers widely attributed this improvement to the system’s design of an agent that considers holistic adjustments rather than isolated modifications.

In our observations, we found that when teachers worked on the baseline task of adapting lesson plans with ChatGPT, they often neglected to adjust the learning profile first. As a result, they had little basis for deciding how to modify the tasks and typically resorted to simple prompts such as “please make the task easier or harder.” However, this is precisely where ChatGPT struggles, as it does not reliably capture task difficulty, leading to weak baseline performance. By contrast, when using AdaPT, teachers were required to go through the learning profile analysis stage as part of the workflow, which in turn gave them a clearer sense of how to adjust the tasks.

\subsubsection{RQ3: Reasonableness of Objectives and Tasks}
The clarity and feasibility of objectives and tasks were rated highly, with averages of 5.83 for clarity of objectives (Q12), 6.33 for time-fit with lesson duration (Q13), and 6.50 for task operability (Q14). The dimension on balancing basic and advanced tasks (Q15) received a mean of 5.33, while still higher than the baseline.

Interview data offered further insights into these findings. Similar to the challenges noted under RQ2, several teachers felt that the system’s transformation brought only limited improvements compared with directly using LLMs, primarily due to large models’ limited grasp of task difficulty. As P6 commented, ``\textit{the descriptions of learning objectives are not always sufficiently clear, and the tasks do not fully capture a rich spectrum of difficulty levels}.'' 
In our observations, we also noted that in many cases, AdaPT explicitly indicated which weaker aspects of the learning profile triggered the system to lower the difficulty of certain tasks. However, when it came to generating specific items, the LLM’s inherent insensitivity to difficulty levels still led to occasional failures. 

Nonetheless, teachers consistently affirmed the practical value of AdaPT-designed lesson plans from the standpoint of workload and implementation. As P5 explained, ``\textit{in terms of the number of tasks and their operability in Yunnan, the system-generated plans seem very good}.'' In addition, we found that some teachers struggled to adjust lesson plans primarily because they found it difficult to integrate appropriate contextual elements into learning tasks. This is precisely where AdaPT demonstrated a strong advantage: by automatically embedding culturally and situationally relevant contexts, the system significantly reduced the creative burden. As a result, teachers who were less skilled at contextual adaptation perceived AdaPT as offering a substantial advantage. This consensus underscores that while there is still room to refine the handling of difficulty gradients, the system already provides considerable benefits for teachers’ day-to-day practice.

\subsubsection{RQ4: Explanations and Traceability}
The system’s provision of adaptation rationales and evidence was one of the most appreciated features. Teachers rated it highly across all items: 6.75 for providing reasons (Q16), 6.58 for grounding in profile evidence (Q17), 6.08 for reasonableness (Q18), and 6.83 for supporting further optimization (Q19). Ratings for RQ4 were still based on a 7-point Likert scale, where 4 indicated a neutral evaluation, 7 represented that AdaPT was extremely helpful, and 1 indicated that it was not helpful at all.

Teachers used two complementary approaches to inspect the transformation results: directly comparing the original and adapted lesson plans (\autoref{fig4}-B1), or examining the evidence presented in the \textit{Review Window} (\autoref{fig4}-C), depending on their preferred workflow. After some time working with the system, all teachers reported that once they became familiar with the visualization lines in the interface, ``\textit{the use of connecting lines to visualize the reasons made the adaptation rationale clear at a glance}.''

\subsubsection{RQ5: Overall Educational Value}
Finally, participants rated the practical value of the adapted lesson plans positively. Scores included 6.25 for teachability (Q20) and 5.67 for learnability (Q21), indicating that teachers generally found the plans both implementable and accessible for students. Interview reflections echoed this perception. P12 emphasized that the AdaPT-transformed plans had ``\textit{a clear structure and complete task design},'' which allowed teachers to directly implement them in class and, at the same time, ``\textit{gain a better understanding of the LLM’s design intentions—an unexpected benefit}.'' P4 further highlighted the system’s accessibility, noting that ``\textit{if deployed in rural areas, its ease of use would make it highly suitable for local teachers, who could achieve good results without needing prior AI expertise}.''

In addition, we have begun piloting the system at the school where P3 and P4 teach. 
Two primary usage scenarios were identified: 
(1) cross-school adaptation, where the school’s lesson plans are shared with partner schools within the same region that require localized adjustments; and (2) within-school adaptation across different classes. 
In scenario (1), we observed that with the support of AdaPT, P3 and P4 were able to complete lesson plan adaptations within 15–20 minutes, whereas other teachers typically spent more than one hour preparing lesson plans for partner schools. 
In classroom practice, P3 reported that once familiar with AdaPT’s visualization-enhanced interface, the LLM-generated content was particularly helpful in broadening her perspective on how to modify lesson plans. 
P4 explicitly noted that in teaching practice, the AdaPT-transformed lesson plans made it significantly easier for students in the partner school to understand and engage with the material.

\begin{figure}[h]
  \centering
  \includegraphics[width=\linewidth]{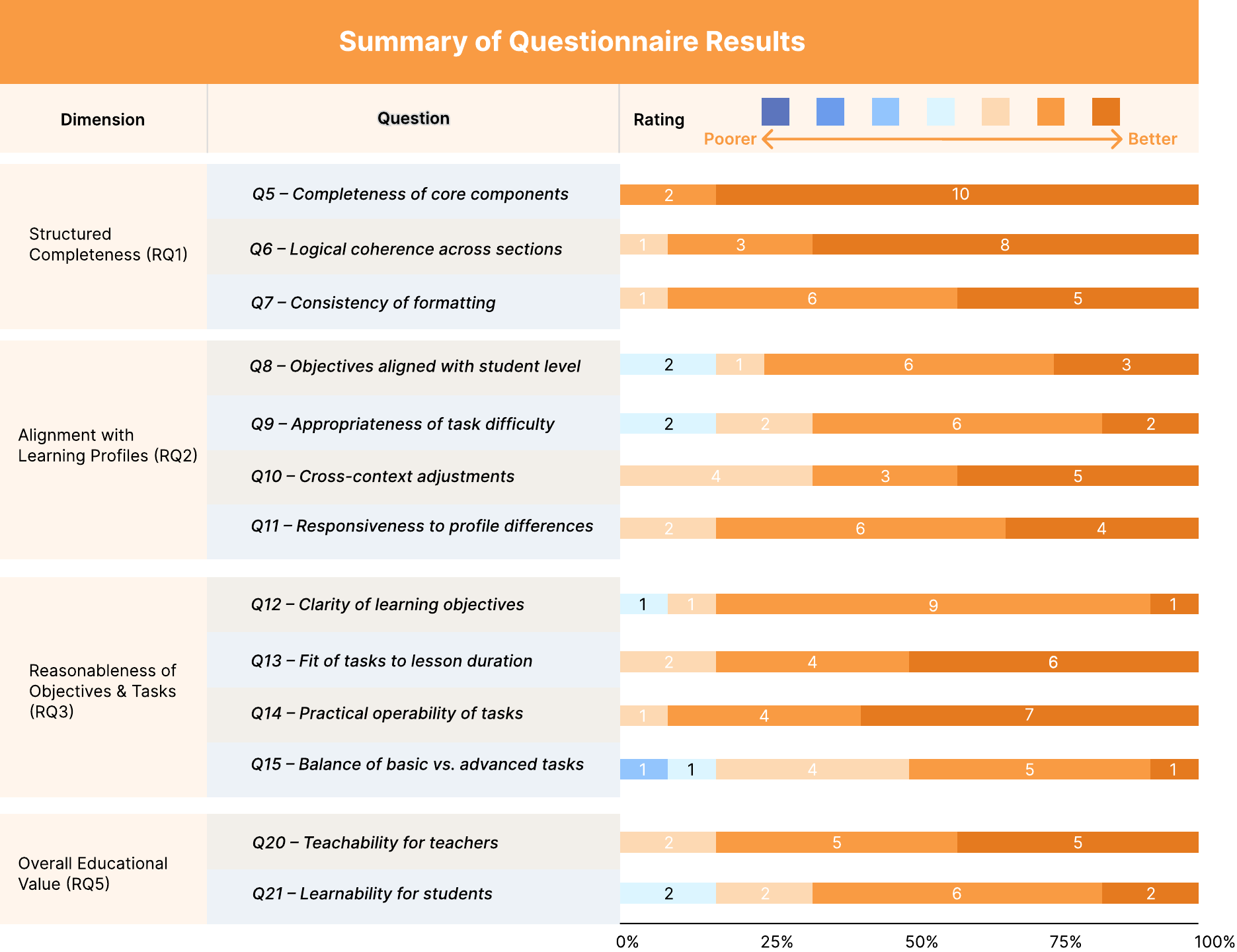}
  \caption{Comparison of questionnaire results between the baseline workflow (using ChatGPT for lesson plan adaptation) and the AdaPT system. The orange blocks represent areas where teachers consider AdaPT performs better than the baseline, and vice versa. Results are summarized across RQ1, RQ2, RQ3, RQ5; note that RQ4 is excluded from the comparison since baseline does not provide reasons or traces.}
  \Description{Bar chart comparing questionnaire ratings for baseline GPT workflow and AdaPT system across RQ1–RQ5, with RQ4 excluded.}
\end{figure}

\subsection{Quantitative Analysis of Questionnaire Reliability}
To assess the reliability of our questionnaire results, we conducted a quantitative analysis of the internal consistency of participants’ ratings across different evaluation dimensions (Q5–Q21). Since our study does not involve ground truth grading values, we focus on the reliability of the Likert-scale responses themselves. Ensuring reliability is important for validating that participants’ evaluations reflect consistent judgments across items rather than random variation.

We first examined the internal consistency of the questionnaire using Cronbach’s $\alpha$, a widely adopted metric for measuring the reliability of Likert-type scales. The overall Cronbach’s $\alpha$ across all items was $0.86$, indicating strong internal consistency. At the subscale level, reliability was high for structural completeness ($\alpha = 0.86$), alignment with learning profiles ($\alpha = 0.76$), and overall educational value ($\alpha = 0.83$). In contrast, the dimension of reasonableness of objectives and tasks (RQ3) showed a lower $\alpha$ of 0.59. This reflects greater diversity in teachers’ judgments, as different participants faced varying student contexts; while the system tended to align task difficulty with the overall profile, it often overlooked gradient differentiation and showed limitations in accurately adjusting difficulty levels. Similarly, the explanations and traceability (RQ4) dimension yielded an $\alpha$ of 0.68, which was largely due to one participant misinterpreting the system’s differentiation output. Aside from this outlier, most teachers found the rationales and evidence reasonable and acceptable.

In addition, we conducted one-sample t-tests and Wilcoxon signed-rank tests to examine whether participants’ ratings were significantly higher than the neutral baseline of $4$. Results showed that all items (Q5–Q21) were rated significantly above baseline (all $p < .01$ after FDR correction), with mean ratings ranging from $5.33$ to $6.83$. This suggests that teachers consistently perceived the system as substantially more effective than neutral expectations.

Taken together, these results confirm that the questionnaire exhibits strong internal reliability and that participants’ evaluations provide robust evidence of the system’s effectiveness across multiple dimensions.

\section{Discussion}
In this section, we discuss three key themes that emerged from our study. First, we situate AdaPT within the broader context of educational inequality in China, highlighting how the system may contribute to addressing disparities across regions, districts, and schools. Second, we emphasize the importance of teacher-centered design, showing how minimizing cognitive burden and preserving teacher autonomy are crucial for real-world adoption. Third, we reflect on the limitations of LLM-based lesson plan transformation and outline future directions for improving sensitivity to task difficulty, efficiency, and cross-domain learning profile analysis.

\subsection{Educational Inequality in China}

The development of AdaPT is situated within the broader context of educational inequality in China. Persistent disparities exist not only between urban and rural regions but also across districts and even within schools of the same city. These differences pose significant challenges for teachers, who must adapt lesson plans to students with widely varying levels of readiness, access to resources, and prior knowledge. By enabling lesson plan transformation, AdaPT aims to support a more equitable redistribution of teaching resources, helping teachers in under-resourced areas access and adapt instructional materials originally designed for more advantaged contexts. In doing so, the system contributes to ongoing discussions on educational equity and the role of digital tools in narrowing structural gaps.

Our exploration also revealed deeper structural barriers that shape educational inequality. In interviews, teachers engaged in cross-regional support programs often attributed difficulties to what they described as teacher “inertia”—a reluctance to learn new technologies or a lack of motivation to modify lesson plans. Education experts, however, emphasized that such inertia is less a matter of individual disposition than a symptom of systemic economic disparities. In better-resourced regions and schools, funding allows for the hiring of administrators and researchers who provide top-down supervision and pedagogical guidance. In contrast, under-resourced schools lack this structural support, leaving teachers with fewer incentives and capacities to innovate.

The pilot deployment of AdaPT highlighted the importance of institutional leadership in overcoming such barriers. Implementation was facilitated by the school principal’s direct involvement and aligned with the school’s broader agenda for AI-powered educational innovation. This institutional support proved critical for ensuring smooth adoption and meaningful use. Together, these findings underscore that addressing educational inequality cannot rely on technological advances alone; rather, progress depends on integrating macro-level factors—such as economic resources and policy frameworks—with digital tools to achieve sustainable change.

\subsection{Teacher-Centered Design and the Learning Burden}

A central insight from our formative study is the importance of designing AI systems that empower, rather than replace, teachers. Participants repeatedly emphasized that the cognitive burden of learning new systems is one of the primary barriers to adoption. For example, T5 noted that teachers are reluctant to engage with AI systems that impose a steep learning curve, as such tools add to their workload instead of reducing it. In contrast, P4 praised AdaPT’s simplicity and strong explanatory UI, arguing that teachers in underdeveloped regions could readily operate the system. Based on his own teaching experiences in rural support programs, he highlighted that systems which ``\textit{let teachers be lazy}'' without adding extra burden are precisely what educators need.

This theme was echoed by P1, who criticized the commercialization of AI-driven education platforms that overlook teachers’ real needs. For instance, general-purpose agent platforms are poorly suited for classrooms because teachers neither understand nor have time to learn AI-related concepts. Instead, designing abstractions that map AI’s domain knowledge into teachers’ practical instructions is key to real-world adoption. 

Similarly, in broader discussions of educational informatization, expert E2 observed that many existing initiatives have focused on replacing rather than empowering teachers. She criticized traditional “One Screen”–style solutions that merely deliver recorded lectures, arguing that such approaches risk further disconnecting instructional content from students’ actual learning needs. Instead, she advocated for blended formats—for example, combining short recordings with teacher-led discussion and group practice—that preserve teacher agency and better align instruction with classroom realities. Together, these findings highlight the design imperative of teacher-centered systems: minimizing cognitive load, preserving teacher autonomy, and providing practical scaffolds that integrate seamlessly into existing pedagogical workflows.

\subsection{Limitations of LLM-Based Lesson Plan Transformation and Future Directions}

Our study surfaced several limitations of using general-purpose large language models for lesson plan transformation.

First, LLMs have limited sensitivity to task difficulty. Although they can generate learning tasks of acceptable quality, they often fail to capture clear gradients between basic and advanced levels. 
This limitation makes it difficult for teachers to rely on the model for fine-grained differentiation of objectives and tasks. 
Exploring how LLMs can better represent and reason about “difficulty” therefore emerges as an intriguing and important direction for future work.

Second, the computational cost is high. To compensate for this lack of difficulty awareness and contextual precision, prompts must be much longer and more detailed, which results in significant token overhead. 
This creates practical barriers for adoption in real-world teaching scenarios where efficiency is essential. 
Future research should examine how to support teachers in generating more effective prompts—without requiring prior AI expertise—so that the system remains accessible while reducing computational overhead.

Third, learning profile analysis is inherently complex. 
Educational outcomes are shaped by a wide range of interdependent factors, many of which go beyond what our current design can capture. 
As E1 pointed out, readiness in one subject often depends on skills from another—for instance, language proficiency can influence students’ ability to solve mathematics word problems. 
Addressing these cross-domain dependencies requires more sophisticated learning analytics approaches. 
We envision future research exploring how to combine data-driven methods from educational data mining and learning analytics with LLM-based reasoning, enabling more holistic and context-aware lesson plan transformations.

\section{Conclusion}
In this paper, we introduced AdaPT, an LLM-powered system that transforms existing lesson plans to support cross-regional and differentiated instruction. Grounded in formative studies with teachers and experts, AdaPT reframes lesson preparation from creation to transformation, aligning with teachers’ natural workflows. Our evaluation demonstrated that AdaPT reduces preparation workload, improves contextual relevance, and preserves teacher agency through transparent rationales and iterative control. This work contributes to advancing human–AI collaboration in education by showing how lesson plan transformation can promote instructional equity while minimizing teachers’ cognitive burden. Future work should enhance learning profile analysis with educational data, expand support for diverse disciplines, and refine collaboration mechanisms to scale AdaPT as a teacher-centered pathway toward equitable educational technologies.


\section*{Acknowledgments}
Sicheng Song, and Minyu Wu are the corresponding authors.

\bibliographystyle{ACM-Reference-Format}
\bibliography{references}

\end{document}